\documentclass[aps,pra,twocolumn,superscriptaddress]{revtex4-2}
\usepackage{amsfonts}
\usepackage{graphicx}
\usepackage{epstopdf}
\usepackage{epsfig}
\usepackage{amsmath}
\usepackage[normalem]{ulem}
\usepackage{multirow}
\usepackage{makecell}
\usepackage{array}
\usepackage{booktabs}
\usepackage{mathtools}
\usepackage{bm}
\usepackage[dvipsnames]{xcolor}
\usepackage{array}
\usepackage{booktabs}
\usepackage{comment}

\usepackage{tabularx}
\newcolumntype{Y}{>{\centering\arraybackslash}X}
\usepackage{blindtext} 

\begin{document}
\title{An experimental scheme for determining the Berry phase in two-dimensional quantum materials with a flat band}

%$\alpha$-$\mathcal{T}_3$ lattice}
%\title{Interplay of ballistic transport and flat band}

\author{Li-Li Ye }
\affiliation{School of Electrical, Computer and Energy Engineering, Arizona State University, Tempe, Arizona 85287, USA}

\author{Cheng-Zhen Wang}
\affiliation{Wave Transport in Complex Systems Lab, Department of Physics, Wesleyan University, Middletown, CT-06459, USA}

\author{Ying-Cheng Lai} \email{Ying-Cheng.Lai@asu.edu}
\affiliation{School of Electrical, Computer and Energy Engineering, Arizona State University, Tempe, Arizona 85287, USA}
\affiliation{Department of Physics, Arizona State University, Tempe, Arizona 85287, USA}

\begin{abstract}

	Experimentally feasible methods to determine the Berry phase, a fundamental quantity characterizing a quantum material, are often needed in applications. We develop an approach to detecting the Berry phase by using a class of two-dimensional (2D) Dirac materials with a flat band, the $\alpha$-$\mathcal{T}_3$ lattices. The properties of this class of quantum materials are controlled by a single parameter $0 \le \alpha \le 1$, where the left and right endpoints correspond to graphene with pseudospin-1/2 and the dice lattice with pseudospin-1 Dirac-Weyl quasiparticles, respectively, and each specific value of $\alpha$ represents a material with a unique Berry phase. Applying a constant electric field to the $\alpha$-$\mathcal{T}_3$ lattice, we calculate the resulting electric current and find a one-to-one correspondence between the current and the Berry phase in both the linear and nonlinear response regimes. In the linear (Kubo) regime, the main physics is the Zitterbewegung effect. In the nonlinear regime, the Schwinger mechanism dominates. Beyond the nonlinear regime, Bloch-Zener oscillations can arise. Measuring the current thus provides an effective and experimentally feasible way to determine the Berry phase for this spectrum of 2D quantum materials.

\end{abstract}
\date{\today}

\maketitle

\section{Introduction} \label{sec:intro}

The Berry phase (the geometric phase or the Pancharatnam-Berry 
phase)~\cite{Pancharatnam:1956,LHOPS:1958,Berry:1984} of the electronic wave 
function is a fundamental characteristic of quantum materials and can have 
significant effects on material properties and physical phenomena such as 
polarization, magnetism, and quantum anomalous spin Hall effects~\cite{XCN:2010}.  
The Berry phase arises when a parameter of the system completes a cycle of 
adiabatic changes: even as the parameter returns to its initial value, the 
wave function gains an extra phase of purely geometric origin. The value of 
the Berry phase depends on the nature of the quasiparticles which, in turn, 
depends on the specific quantum material. Given a family of quantum materials,
the Berry phase is effectively a unique identifier of each material in the family. 
For example, monolayer graphene hosting a pair of Dirac cones and pseudospin-1/2 
quasiparticles, the Berry phase is $\pm\pi$ associated, respectively, with the 
electronic states in the two Dirac cones~\cite{ZTSK:2005,CU:2008}. For bilayer 
graphene, the Berry phase is $2\pi$, which leads to unconventional quantum Hall 
effect~\cite{NMMFKZJSG:2006}. For pseudospin-1 Dirac-Weyl materials, the Berry 
phase is zero~\cite{ICN:2015,WHXL:2019}. In recent years, various two-dimensional 
(2D) Dirac materials have been discovered at a rapid 
pace~\cite{GN:2007,GG:2013,AKB:2016}, each carrying a unique type of quasiparticles
with a unique value of the Berry phase. Given a new quantum material, knowing the 
Berry phase is thus of theoretical, experimental, and applied interests.

In principle, the phenomenon of Aharonov-Bohm interference provides an approach
to assessing the Berry phase~\cite{DLRBSS:2015,RL:2016,Ghaharietal:2017}. Take
graphene as an example. For a circular graphene $p$-$n$ junction resonator, 
due to the $\pm\pi$ Berry phase of the quasiparticles, as the strength of an
external magnetic field is tuned, a sudden change in the energy of the 
angular-momentum states can occur, providing an indirect way to ascertain the 
value of the Berry phase~\cite{RL:2016,Ghaharietal:2017}. For photonic crystals, 
their analogy with graphene was exploited to measure the geometric Berry 
phase by removing the dynamical phase~\cite{SNB:2008}. For a general family of
2D Dirac-Weyl materials (the $\alpha$-$\mathcal{T}_3$ lattices), the semiclassical
dynamics of a chaotic cavity made of such a material were explored to infer its
Berry phase~\cite{WHXL:2019}. In particular, by applying a gate voltage to generate
a quasi-confinement of a certain geometric shape that generates chaos in the classical
limit, a one-to-one correspondence between the exponential rate of particles 
escaping from the cavity and the Berry phase was identified. Despite the 
theoretical appeal of this semiclassical phenomenon, experimentally monitoring the 
decay of an ensemble of quasiparticles from a cavity of certain quantum material is
not feasible at present.

In this paper, we present theoretical calculations leading to an experimentally 
feasible approach to detecting the Berry phase for the $\alpha$-$\mathcal{T}_3$ 
lattice family whose material properties are controlled by a single parameter: 
$0 \le \alpha \le 1$. An $\alpha$-$\mathcal{T}_3$ lattice has the honeycomb lattice
as its base with an additional atom at the center of each hexagonal unit cell. In 
the tight-binding approximation, the center atom couples with any of the hexagonal 
atoms with the energy $\alpha t_{\epsilon}$, where $t_{\epsilon}$ is the 
nearest-neighbor coupling energy of the honeycomb lattice. Because of the center 
atom, an $\alpha$-$\mathcal{T}_3$ lattice with $\alpha > 0$ possesses three distinct 
energy bands: a pair of Dirac cones and a flat band through the contact point of the 
two Dirac cones. As $\alpha$ increases from zero, a continuous spectrum of 2D 
Dirac-Weyl materials is generated: from the graphene ($\alpha=0$) to the dice lattice 
($\alpha=1$), and the corresponding Berry phase can change from $\pi$ to zero. As a 
result of the continuous decrease in the Berry phase, a number of pertinent physical 
phenomena change their characteristics. For example, the flat band plays an important role in the accurate quantization of the Hall conductivity in the case of an $\alpha$-$\mathcal{T}_3$  lattice~\cite{biswas:2016}. Wave packet dynamics and the Zitterbewegung effect~\cite{biswas:2018}, the effect of periodic kicks~\cite{tamang:2021}, Floquet dynamics driven by the electromagnetic field~\cite{dey:2018}, Size effects on atomic collapse~\cite{oriekhov:2023}, and topological phase transitions~\cite{tamang:2023} have also been explored in an $\alpha$-$\mathcal{T}_3$ lattice. The orbital magnetic 
response~\cite{Raoux:2014} at the Dirac point changes from diamagnetic ($\alpha=0$) 
to paramagnetic ($\alpha=1$) and the nature of the Hall quantization~\cite{ICN:2015} 
switches from relativistic to nonrelativistic. Moreover, the patterns of optical 
response~\cite{han:2022} and magneto-optical modulation~\cite{chen:2019} change 
because they depend on the interband transitions among the three bands. Further, 
optical conductivity quantization and higher-order harmonic generation were 
observed~\cite{chen2019nonlinear}, so was the effect of a broken flat-band on the 
integer quantum Hall effect by the disorder or staggered lattice 
potential~\cite{wang2020integer}. Experimentally, the $\alpha$-$\mathcal{T}_3$ 
lattice has been realized in the critical doping material~\cite{malcolm:2015}
$\textnormal{Hg}_{1-\textnormal{x}}\textnormal{Cd}_{\textnormal{x}}\textnormal{Te}$.
For $\alpha=1$, the dice lattice described by the pseudospin-$1$ Dirac-Weyl
Hamiltonian can be grown in the transition-metal oxide~\cite{wang:2011}
$\textnormal{SrTiO}_3$/$\textnormal{SrIrO}_3$/$\textnormal{SrTiO}_3$ or in
graphene-$\textnormal{In}_{2}\textnormal{Te}_{2}$ bilayer~\cite{giovannetti:2015}.

We focus our calculations on the electric current density (or simply the current) produced when 
a constant electric field is applied to the $\alpha$-$\mathcal{T}_3$ lattice. In the 
classical Drude picture, when driven by a constant electric field $\mathcal{E}$, the momentum of 
the electrons in ballistic transport will increase with time: $q=e\mathcal{E}t$. 
Nevertheless, Dirac electrons will be excited instantaneously to the Fermi velocity 
(pinned to the ``light cone'')~\cite{dora:2010}, where the excitation mechanism is 
described by the Schwinger effect~\cite{schwinger:1951} or the Landau-Zener 
dynamics~\cite{landau:1932,zener:1932} that occur where there are two 
avoided-crossing energy levels under the adiabatic evolution induced by the electric 
field. Another relevant phenomenon is Bloch oscillations~\cite{bloch:1928,zener:1934} 
in the time evolution of the electronic states in a single energy band. When multiple 
bands without crossings are present, Bloch-Zener 
oscillations~\cite{dreisow:2009,lim:2012,zhang:2021} can take place. For the 
$\alpha$-$\mathcal{T}_3$ lattice, irregular Bloch-Zener oscillations~\cite{Ye:2023} 
can arise, due to the mixed interference of the quantum states in multi-bands based on Landau-Zener-Stuckelberg-Majorana transitions~\cite{ivakhnenko:2023,kofman:2023}. In 
addition, the mass term associated with the Dirac electrons or a weak disorder can 
render a nonzero minimal conductivity that depends on the value of 
$\alpha$~\cite{wang:2020}. Other relevant transport phenomena in the 
$\alpha$-$\mathcal{T}_3$ lattice includes the linear response in graphene with the 
chiral anomaly and nonlinear response when the perturbation theory breaks 
down~\cite{kao:2010}, as well as nonlinear conductivity with THz-induced charge 
transport~\cite{sato:2021}. Nonequilibrium dynamics beyond the linear response in 
3D Weyl semimetals~\cite{vajna:2015} and nodal loop semimetals~\cite{okvatovity:2021} 
have also been studied.

The main physical considerations behind our calculations of the current are as 
follows. On different time scales, the transport properties and the physical 
mechanisms are distinct. In particular, in the Kubo regime~\cite{dora:2010} under 
the weak field approximation, the average current density is saturated and dominated 
by the Zitterbewegung effect~\cite{katsnelson:2006} originated from the interference 
between the energy bands, which defines the regime of linear response. A strong 
electric field places the system in the Schwinger regime~\cite{schwinger:1951}, 
where the electrons are excited by the Schwinger mechanism in which the vacuum field 
loses energy to produce electron-positron pairs. The transition probabilities among 
the energy bands are described by the Landau-Zener dynamics~\cite{vitanov:1996}, 
where the quasiparticles adiabatically evolve and transitions occur about 
the point at which the two levels are closest to each other but without crossing. 
In this regime, the current is proportional to the number of the excited particles, 
representing a nonlinear response. When the product of the electric field and time is
comparable to the lattice constant, Bloch oscillations~\cite{bloch:1928,zener:1934} 
become important in the Landau-Zener dynamics, leading to Bloch-Zener 
oscillations~\cite{dreisow:2009,lim:2012,zhang:2021}. The main finding is a 
monotonic dependence of the current on the materials parameter $\alpha$ in both the 
linear and nonlinear response regimes, implying a one-to-one correspondence between 
the current and the Berry phase, thereby providing a possible experimental scheme 
to determine the latter.

Our main code is uploaded to GitHub:
https://github.com/liliyequantum/Berry-phase.

\section{Current and Berry phase calculation for $\alpha$-$\mathcal{T}_3$ lattice}

\subsection{Zero-field effective Hamiltonian} \label{subsec:zeros_field_alpha_T3_model}

\begin{figure} [ht!]
\centering
\includegraphics[width=\linewidth]{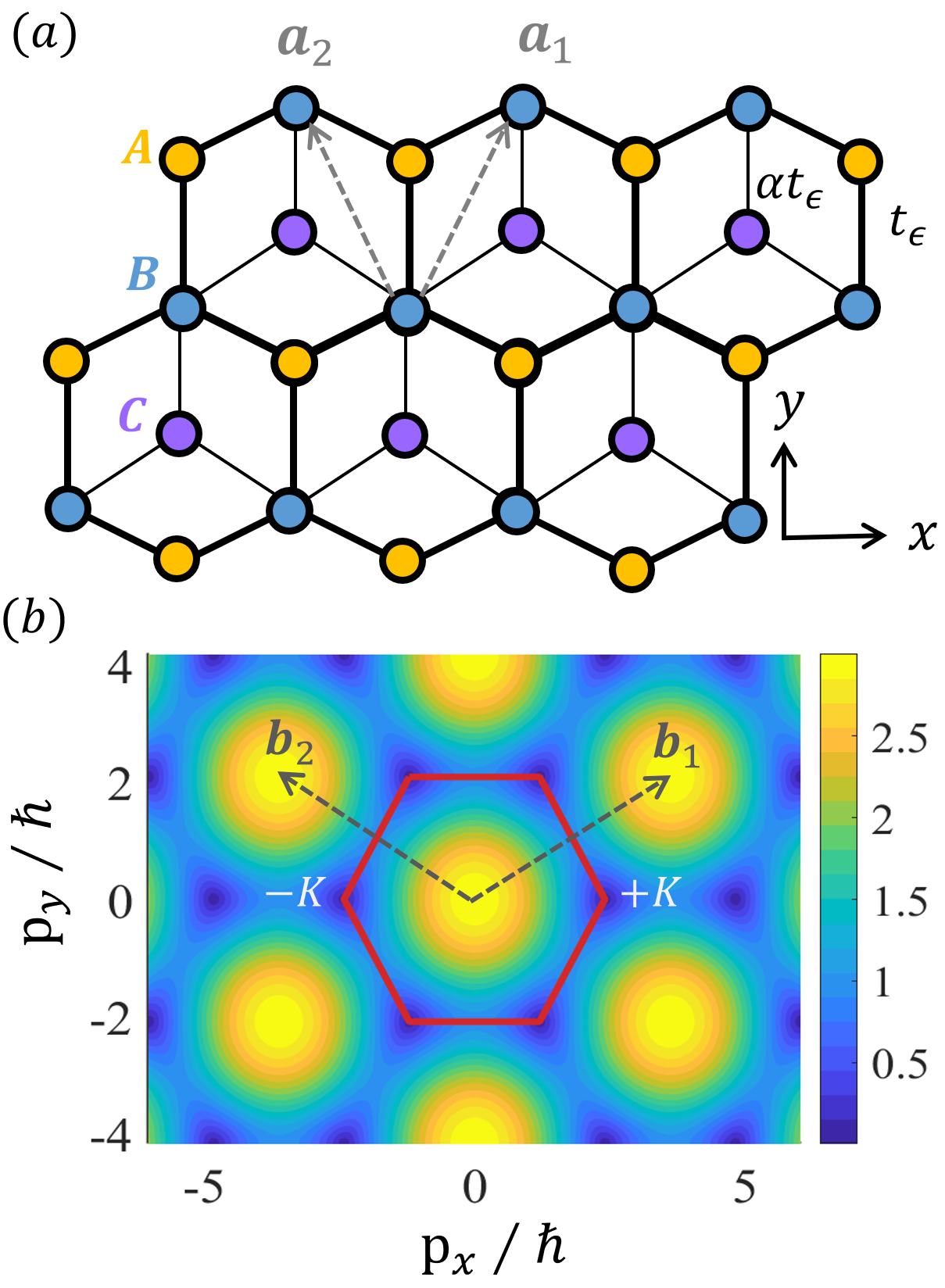}
\caption{Structure of $\alpha$-$\mathcal{T}_3$ lattice and zero field energy-band 
structure. (a) The lattice structure, where each unit cell contains three distinct
atoms. The nearest-neighbor hopping energy between the A and B sites is 
$t_{\epsilon}$ and that between B and C sites is $\alpha t_{\epsilon}$. The material 
parameter $\alpha$ characterizes the relative coupling strength between the flat band 
and the Dirac-cone bands. (b) The zero-field energy spectrum of the positive dispersion 
band in the 2D momentum space for any $\alpha$ value, where the red hexagon denotes 
the first Brillouin zone and the zero energy points give two topologically 
nonequivalent Dirac points $\zeta \mathbf{K}$ with the valley index $\zeta=\pm1$.}
\label{fig:alpha_T3_lattice}
\end{figure}

To calculate the current and the Berry phases in an $\alpha$-$\mathcal{T}_3$ lattice, 
we begin with the zero-field lattice Hamiltonian. The basic lattice structure is 
shown in Fig.~\ref{fig:alpha_T3_lattice} (a), where there are three distinct atoms in 
a unit cell: A and B atoms belonging to the base hexagonal lattice, and C atom at 
the center of the unit cell. The tight-binding Hamiltonian is given 
by~\cite{Raoux:2014}
\begin{align}
    H=\left[\begin{array}{ccc}
0 & f_{\mathbf p}\cos\varphi & 0\\
f_{\mathbf p}^{*}\cos\varphi & 0 & f_{\mathbf p}\sin\varphi\\
0 & f_{\mathbf p}^{*}\sin\varphi & 0
\end{array}\right],
\end{align}
with
\begin{equation}\label{eq:f_p}
	f_{\mathbf p}=-t_{\epsilon}\left(1+e^{-i{\mathbf p}\ensuremath{\cdot}\mathbf{a}_{1}/\hbar}+e^{-i{\mathbf p} \cdot {\mathbf a}_{2}/\hbar}\right),
\end{equation}
where ${\mathbf p}=(p_{x},p_{y})$ and $t_{\epsilon}$ is the nearest-neighbor hopping energy between an A and a B atom with the parametrization~\cite{Raoux:2014}: 
$\tan\varphi=\alpha\in[0,1]$. In the position space, the primitive translation 
vectors are ${\mathbf a}_{1}=a\left(\sqrt{3}/2,3/2\right)$, 
$\mathbf{a}_{2}=a\left(-\sqrt{3}/2,3/2\right)$,
where $a$ is the lattice constant (inter-site distance). The base vectors in the 
reciprocal lattice of the hexagonal Brillouin zone are
${\mathbf b}_{1}=\left(\sqrt{3}/3,1/3\right)2\pi/a$ and 
${\mathbf b}_{2}=\left(-\sqrt{3}/3,1/3\right)2\pi/a$. The
eigenenergy spectrum~\cite{Raoux:2014} of $\alpha$-$\mathcal{T}_3$  lattice is 
independent of the value of $\alpha$, which consists of a zero energy flat band 
$E_{0}=0$ and two linearly dispersive bands $E_{\lambda}=\lambda\left|f_{\mathbf p}\right|$ 
with the band index $\lambda=\pm$. The structure of the positive band is shown in 
Fig.~\ref{fig:alpha_T3_lattice}(b). There are two nonequivalent Dirac contact 
points: $+\mathbf{K}=\left(2/(3\sqrt{3}),0\right)2\pi/a$ and 
$-\mathbf{K}=\left(-2/(3\sqrt{3}),0\right)2\pi/a$. 

Denoting the momentum vector from a Dirac point $\zeta \mathbf{K}$ as $\mathbf{q}$ and linearizing the 
corresponding function $f_{\mathbf q}$ in the Hamiltonian as
\begin{equation}\label{eq:f_q}
 f_{\mathbf q}\approx  v_{F}\left(\zeta q_{x}-iq_{y}\right),
\end{equation}
where $v_F$ is Fermi velocity, $q_x,\;q_y$ are the momentum components measured from Dirac points with the valley index $\zeta=\pm1$, we 
obtain the effective Hamiltonian in the continuum limit at low energy excitation as
\begin{align} \label{eq:zero_field_effective_H}
     H_{\mathbf q}\approx  v_{F}\left(\zeta S_{x}^{\prime}\left(\varphi\right)q_{x}+S_{y}^{\prime}\left(\varphi\right)q_{y}\right),
\end{align}
with 
\begin{align}
   S_{x}^{\prime}\left(\varphi\right)&\equiv\left[\begin{array}{ccc}
0 & \cos\varphi & 0\\
\cos\varphi & 0 & \sin\varphi\\
0 & \sin\varphi & 0
\end{array}\right],\\
 S_{y}^{\prime}\left(\varphi\right)&\equiv\left[\begin{array}{ccc}
0 & -i\cos\varphi & 0\\
i\cos\varphi & 0 & -i\sin\varphi\\
0 & i\sin\varphi & 0
\end{array}\right].
\end{align}
For $\alpha=0$, $S^{\prime}_{x}(0)$ and $S^{\prime}_{y}(0)$ become 
$\sigma_{x}\oplus0$ and $\sigma_{y}\oplus0$, respectively, leading to
\begin{align}
    H_{\mathbf q}|_{\alpha=0}&= v_{F}\left(\zeta\sigma_{x}q_{x}+\sigma_{y}q_{y}\right)\oplus0,\\
    H_{{\rm Spin}-\frac{1}{2}}&= v_{F}\left(\sigma_{x}q_{x}+\sigma_{y}q_{y}\right).
\end{align}
For $\alpha=1$, $S^{\prime}_{x}(\pi/4)$ and $S^{\prime}_{y}(\pi/4)$ are $S_{x}$ and 
$S_{y}$, respectively, i.e., the components of the spin-1 matrix vector. In this case,
we have
\begin{align}
    H_{\mathbf q}|_{\alpha=1}&= v_{F}\left(\zeta S_{x}q_{x}+S_{y}q_{y}\right),\\
H_{\mbox{spin}-1}&=v_{F}\left(S_{x}q_{x}+S_{y}q_{y}\right).
\end{align}
The continuum effective Hamiltonian of the $\alpha$-$\mathcal{T}_3$ lattice, as given
by Eq.~\eqref{eq:zero_field_effective_H}, is a general model that includes the 
pseudospin-$1/2$ and pseudospin-$1$ lattices as the two opposite limiting cases. By 
varying the coupling strength $\alpha\in[0,1]$, a continuous spectrum of Dirac-Weyl
materials with a flat band can be generated.

\subsection{Low-excitation continuum effective $\alpha$-$\mathcal{T}_3$ Hamiltonian in a constant electric field} \label{subsec:continuum_alpha_T3}

We apply a uniform and constant electric field to an $\alpha$-$\mathcal{T}_3$ 
lattice in the $x$ direction starting at time $t=0$, represented by a time-dependent vector potential~\cite{dora:2010,wang:2017}. The corresponding continuum effective 
Hamiltonian around the two nonequivalent Dirac points becomes
\begin{equation}\label{eq:effective_model}
    H_{\mathbf q}\left(t\right)= v_{F}\left(\zeta S_{x}^{\prime}\left(\varphi\right)q_x\left(t\right)+S_{y}^{\prime}\left(\varphi\right)q_{y}\right),
\end{equation}
with $q_x(t)\equiv q_{x}-e\mathcal{A}(t)$, where 
$\mathcal{A}\left(t\right)=\mathcal{E}t\Theta\left(t\right)$ and $\Theta(t)$ is a 
unit step function of time. The quantum dynamics are governed by
\begin{align}
    i\hbar\partial_{t}\psi_{\mathbf q}\left(t\right)=H_{\mathbf q}\left(t\right)\psi_{\mathbf q}\left(t\right).
\end{align}
In the Landau-Zener adiabatic basis~\cite{dora:2010,vitanov:1996}, under an 
infinitesimal electric field, the evolution of a quantum state is described by
\begin{align}
    U^{\dagger}_{\mathbf q}\left(t\right)H_{\mathbf q}\left(t\right)U_{\mathbf q}\left(t\right)=S_{z}\varepsilon_{\mathbf q}\left(t\right),
\end{align}
where $S_z$ is the $z$-component of the spin-1 matrix vector and $U_{\mathbf q}(t)$ 
is given by
\begin{align}\label{eq:unitary_transformation}
   \left[\begin{array}{ccc}
\frac{1}{\sqrt{2}}\cos\varphi\;e^{i\theta_{\mathbf q}} & \sin\varphi\;e^{i
\theta_{\mathbf q}} & \frac{1}{\sqrt{2}}\cos\varphi\;e^{i\theta_{\mathbf q}}\\
\frac{1}{\sqrt{2}} & 0 & -\frac{1}{\sqrt{2}}\\
\frac{1}{\sqrt{2}}\sin\varphi\;e^{-i\theta_{\mathbf q}} & -\cos\varphi\;e^{-i\theta_{\mathbf q}} & \frac{1}{\sqrt{2}}\sin\varphi\;e^{-i\theta_{\mathbf q}}
\end{array}\right],
\end{align}
with $\theta_{\mathbf q}(t)$ being the phase of $f_{\mathbf q}(t)$ and 
\begin{align} \nonumber
\tan\theta_{\mathbf q}\left(t\right)=-\zeta q_{y}/\left[q_{x}-e\mathcal{A}\left(t\right)\right]
\end{align}
because of
\begin{equation}
    f_{\mathbf q}(t)\approx  v_{F}\left(\zeta q_{x}(t)-iq_{y}\right).
\end{equation}
The eigenstates of positive, zero, and negative energy bands can then be written down
and be distinguished. For example, the positive eigenenergy spectrum 
$\epsilon_{\mathbf q}(t)\equiv+|f_{\mathbf q}(t)|$ is given by
\begin{align}
    \varepsilon_{\mathbf q}\left(t\right)= v_{F}\sqrt{\left(q_{x}-e\mathcal{A}\left(t\right)\right)^{2}+q_{y}^{2}}.
\end{align}
The transformed time-dependent Dirac equation becomes 
\begin{align}\label{eq:dirac_q}
i\hbar&\partial_{t}\Phi_{\mathbf q}\left(t\right)=\left[S_{z}\varepsilon_{\mathbf q}\left(t\right)-\widetilde{S}_{x}\frac{\hbar v_{F}^{2}q_{y}e\mathcal{E}}{\zeta\varepsilon^2_{\mathbf q}\left(t\right)}\right]\Phi_{\mathbf q}\left(t\right),
\end{align}
where 
\begin{align} \nonumber
	\Phi_{\mathbf q}\left(t\right) &\equiv U^{\dagger}_{\mathbf q}\left(t\right)\psi_{\mathbf q}\left(t\right), \\ \nonumber 
	\widetilde{S}_{x} &\equiv S_{x}\sin2\varphi-S_{L}\cos2\varphi, 
\end{align}
with $S_{L}$ defined by
\begin{align}
    S_{L}&\equiv \left[\begin{array}{ccc}
-1/2 & 0 & -1/2\\
0 & 1 & 0\\
-1/2 & 0 & -1/2
\end{array}\right].
\end{align}
The second term in Eq.~\eqref{eq:dirac_q} arises from the time dependence of the 
unitary transformation 
$-i\hbar U^{\dagger}_{\mathbf q}(t)\partial_{t}U_{\mathbf q}(t)$. Consider 
the initial state in which the lower Dirac cone is fully occupied:
\begin{align}
   \Phi_{\mathbf q}\left(t=0\right)=\left[0,0,1\right]^{T},
\end{align}
the average current density $\langle \mathcal{J}_{x}\rangle_{\mathbf q}\left(t\right)$
in the momentum space is invariant under the unitary transformation: 
\begin{align}\label{eq:current_original}
    \langle \mathcal{J}_{x}\rangle_{\mathbf q}\left(t\right)&\equiv-e[\psi_{\mathbf q}\left(t\right)]^{\dagger}\left(\partial_{q_{x(t)}}H_{\mathbf q}\left(t\right)\right)\psi_{\mathbf q}\left(t\right),\\ \nonumber
    &=-e v_{F}\zeta\;
	\Phi^{\dagger}_{\mathbf q}\left(t\right)\left[U^{\dagger}_{\mathbf q}\left(t\right)S'_{x}\left(\varphi\right)U_{\mathbf q}\left(t\right)\right]\Phi_{\mathbf q}\left(t\right).
\end{align}
In the adiabatic basis, $\Phi_{\mathbf q}\left(t\right)$ can be expressed as
\begin{align}
   \Phi_{\mathbf q}\left(t\right)=\left[\begin{array}{ccc}
\xi_{\mathbf q}\left(t\right), & \gamma_{\mathbf q}\left(t\right), & \beta_{\mathbf q}\left(t\right)\end{array}\right]^{T},
\end{align}
where $|\xi_{\mathbf q}|^2$, $|\gamma_{\mathbf q}|^2$, and $|\beta_{\mathbf q}|^2$ 
are the probabilities of finding the quasiparticle in the upper, flat, and lower band,
respectively. The average current density can be decomposed into two 
parts~\cite{dora:2010},
\begin{align}
  \langle \mathcal{J}_{x}\rangle_{\mathbf q}\left(t\right)=\langle \mathcal{J}_{x}\rangle_{\mathbf q}^{intra}\left(t\right)+\langle \mathcal{J}_{x}\rangle_{\mathbf q}^{inter}\left(t\right),
\end{align}
which are the intraband and interband currents, respectively, given by 
\begin{widetext}
\begin{align} \label{eq:continuum_current density_1}
\langle \mathcal{J}_{x}&\rangle_{\mathbf q}^{intra}\left(t\right)=-ev_{F}\zeta\cos[\theta_{\mathbf q}\left(t\right)]\left(\left|\xi_{\mathbf q}\left(t\right)\right|^{2}-\left|\beta_{\mathbf q}\left(t\right)\right|^{2}\right), \\ \label{eq:continuum_current density_2}
\langle \mathcal{J}_{x}&\rangle_{\mathbf q}^{inter}\left(t\right)=-e v_{F}\zeta\sin[\theta_{\mathbf q}\left(t\right)]\biggl(2\cos[2\varphi]\;\Re\left[i\xi^{*}_{\mathbf q}\left(t\right)\beta_{\mathbf q}\left(t\right)\right] + \sqrt{2}\sin[2\varphi]\;\Re\left[i\xi^{*}_{\mathbf q}\left(t\right)\gamma_{\mathbf q}\left(t\right)+i\gamma^{*}_{\mathbf q}\left(t\right)\beta_{\mathbf q}\left(t\right)\right]\biggr),
\end{align}
\end{widetext}
with 
\begin{align} \nonumber
	\sin[2\varphi] &=2\alpha/(1+\alpha^{2}), \\ \nonumber
	\cos[2\varphi] &=(1-\alpha^{2})/(1+\alpha^{2}). 
\end{align}
The intraband component represents the current density of the electrons and holes 
in the upper and lower band, respectively, with the opposite signs. The interband 
component depicts the current density due to the interference between the upper, 
flat and lower bands, where the material parameter $\alpha$ modulates contributions 
to the current density from the coupling between energy bands. In particular, for 
$\alpha=0$, the only contribution to the current is the transition from the lower 
to the upper band. However, for $\alpha=1$, the current density is due to the 
coupling between the flat band and the other bands. For $\alpha\in(0,1)$, the 
interband current density is a mixture of the two extreme cases.

Using the normalization condition, we have 
\begin{equation} \label{eq:normalization}
    \left|\xi_{\mathbf q}\left(t\right)\right|^{2} - \left|\beta_{\mathbf q}\left(t\right)\right|^{2} = 2\left|\xi_{\mathbf q}\left(t\right)\right|^{2}+\left|\gamma_{\mathbf q}\left(t\right)\right|^{2}-1.
\end{equation}
Substituting Eq.~\eqref{eq:normalization} into the intraband current in 
Eq.~\eqref{eq:continuum_current density_1}, the constant in the third term of 
Eq.~\eqref{eq:normalization} will vanish~\cite{dora:2010} after a momentum integration. The integration region is theoretically infinite over the momentum space from one $\zeta \pmb{\mbox{K}}$ point, and the integrand is the periodic function $\cos[\theta_{\pmb{q}}(t)]$ from Eq.~\eqref{eq:continuum_current density_1}. Generally, the periodic term, when subjected to integration over an infinite region, is expected to vanish. As a result, the intraband current can be regarded 
as the contribution from the flat and upper bands. Specifically, we denote
$\mathcal{J}^{\textnormal{intra}}(t)$ and $\mathcal{J}^{\textnormal{inter}}(t)$ as 
the momentum integration of 
$\langle \mathcal{J}_x\rangle_{\mathbf q}^{\textnormal{intra}}(t)$,
and $\langle \mathcal{J}_x\rangle_{\mathbf q}^{\textnormal{inter}}(t)$, respectively. 
It is not realistic to numerically integrate over the infinite momentum space from one $\zeta \pmb{\mbox{K}}$ point. Therefore, the integration region for the continuum model is limited to a finite momentum region to ensure that the current converges with
\begin{equation}
    \mathcal{J}(t) = \mathcal{J}^{\textnormal{intra}}(t) + \mathcal{J}^{\textnormal{inter}}(t).
\end{equation}
Note that the effective model described by Eq.~\eqref{eq:effective_model} is derived using a Taylor expansion around a single Dirac point, denoted by $\zeta \pmb{\mbox{K}}$. Therefore, the integration of the average current density from this effective Hamiltonian includes contributions only from this one Dirac point. We can decompose the contributions from all the energy bands in the intraband and 
interband current by integrating Eqs.~\eqref{eq:continuum_current density_1} 
and \eqref{eq:continuum_current density_2} over the 
momentum space and following the term order in
Eqs.~\eqref{eq:continuum_current density_1} 
and \eqref{eq:continuum_current density_2} to define
\begin{align}
    \mathcal{J}^{\textnormal{intra}}(t) &= \mathcal{J}^{\textnormal{intra}}_{\xi}(t) + \mathcal{J}^{\textnormal{intra}}_{\gamma}(t),\nonumber\\
    \mathcal{J}^{\textnormal{inter}}(t) &= \mathcal{J}^{\textnormal{inter}}_{\xi\beta}(t) + \mathcal{J}^{\textnormal{inter}}_{\xi\gamma}(t) + \mathcal{J}^{\textnormal{inter}}_{\gamma\beta}(t).
\end{align}
These expressions are convenient for treating the contributions to the current by
the multiple energy bands in the weak field (Sec.~\ref{subsec:weak_field_regime}) 
and strong field (Sec.~\ref{subsec:strong_field_regime}) cases.

To streamline numerical calculations, we define a number of dimensionless physical 
quantities in the continuum effective $\alpha$-$\mathcal{T}_3$ model:
\begin{align}
    \widetilde{t} &= t/t_0,\nonumber\\
    \widetilde{q}_x &= q_x/q_0,\nonumber\\
    \widetilde{q}_y &= q_y/q_0,\nonumber\\
    \widetilde{\mathcal{E}} &= \mathcal{E}/\mathcal{E}_{0},\nonumber\\
     \widetilde{\varepsilon}_{q}(t) &= \varepsilon_{q}(t)/\varepsilon_0,\nonumber\\
    \widetilde{\mathcal{J}}(t) &= \mathcal{J}(t)/\mathcal{J}_{0},\nonumber\\
     \langle \widetilde{\mathcal{J}}_x\rangle_{q}(t)
     &=\langle \mathcal{J}_x\rangle_{\mathbf q}(t)/\langle \mathcal{J}_{0}\rangle_{\mathbf q},
\end{align}
where $t_0 \equiv \hbar/t_{\epsilon}$, $q_0 \equiv t_{\epsilon}/v_{F}$, 
$\mathcal{E}_{0} \equiv t^{2}_{\epsilon}/(e\hbar v_{F})$, 
$\varepsilon_0 \equiv t_{\epsilon}$, 
\begin{align} \nonumber
	\mathcal{J}_{0}\equiv e^2 \mathcal{E}_0/\hbar\sim e v_{F}/a^2, 
\end{align}
and $\langle \mathcal{J}_{0}\rangle_{\mathbf q} \equiv e v_{F}$.

\subsection{General $\alpha$-$\mathcal{T}_3$ lattice Hamiltonian in a constant electric field} \label{subsec:alpha-T3_lattice_Hamiltonian_and_current density}

With a constant electric field switched on at $t=0$ in the $x$-direction, the 
$x$-component of the momentum is $p_{x}(t)\equiv p_{x}-eEt$. The general Hamiltonian 
of the $\alpha$-$\mathcal{T}_3$ lattice is given by
\begin{align}
   H\left(t\right)=\left[\begin{array}{ccc}
0 & f_{\mathbf p}\left(t\right)\cos\varphi & 0\\
f_{\mathbf p}^{*}\left(t\right)\cos\varphi & 0 & f_{\mathbf p}\left(t\right)\sin\varphi\\
0 & f_{\mathbf p}^{*}\left(t\right)\sin\varphi & 0
\end{array}\right],
\end{align}
where
\begin{equation}\label{eq:f_k_t_lattice}
    f_{\mathbf p}(t)=-t_{\epsilon}\left(1+2\exp{\left(-i\frac{3}{2}\frac{p_{y}a}{\hbar}\right)}\cos\left(\frac{\text{\ensuremath{\sqrt{3}}}}{2}\frac{p_{x}\left(t\right)a}{\hbar}\right)\right).
\end{equation}
The eigenenergy spectrum of the flat band is $\varepsilon_0=0$, and the positive dispersion band in the whole hexagonal 
Brillouin zone is determined by 
$\varepsilon_{\mathbf p}\left(t\right)=+\left|f_{\mathbf p}\left(t\right)\right|$: 
\begin{align}\label{eq:energy}
    \varepsilon_{\mathbf p}(t)=t_{\epsilon}\sqrt{1+4\cos X_{\mathbf p}\left(t\right)\left(\cos Y_{\mathbf p}+\cos X_{\mathbf p}\left(t\right)\right)},
\end{align}
where $X_{\mathbf p}\left(t\right)=\sqrt{3}p_{x}(t)a/(2\hbar)$ and 
$Y_{\mathbf p}=3p_{y}a/(2\hbar)$. The unitary transformation $U_{\mathbf p}(t)$ is 
similar to that in Eq.~(\ref{eq:unitary_transformation}) except that \
$\theta_{\mathbf q}(t)$ is now replaced by $\theta_{\mathbf p}(t)$, which is the 
phase of $f_{\mathbf p}(t)$ in Eq.~\eqref{eq:f_k_t_lattice}. 
The transformed quantum dynamics are governed by~\cite{Ye:2023}
\begin{align}\label{eq:dirac_p}
   i\hbar\partial_{t}\Phi_{\mathbf p}\left(t\right)=\biggl[S_{z}\varepsilon_{\mathbf p}\left(t\right)- \widetilde{S}_{x}\frac{at_{\epsilon}^{2}eE}{\varepsilon_{\mathbf p}^{2}\left(t\right)}C_{\mathbf p}\left(t\right)\biggr]\Phi_{\mathbf p}\left(t\right),
\end{align}
with the coefficient given by
\begin{align} \nonumber
C_{\mathbf p}\left(t\right)=\sqrt{3}\sin Y_{\mathbf p}\sin X_{\mathbf p}\left(t\right). 
\end{align}
The average current density $\langle J_{x}\rangle_{\mathbf p}\left(t\right)$ contains 
two contributions: interband and intraband transitions~\cite{dora:2010}, which can 
generally be written as~\cite{Ye:2023}
\begin{widetext}
\begin{align}\label{eq:lattice_current density_1}
	&\langle J_{x}\rangle_{\mathbf p}^{intra}\left(t\right)=J_{x,\;{\mathbf p}}^{11}\left(t\right)\left(\left|\xi_{\mathbf p}\left(t\right)\right|^{2}-\left|\beta_{\mathbf p}\left(t\right)\right|^{2}\right), \\ \label{eq:lattice_current density_2} 
	&\langle J_{x}\rangle_{\mathbf p}^{inter}\left(t\right)=2\Re\left[J_{x,\;{\mathbf p}}^{13}\left(t\right)\xi_{\mathbf p}^{*}\left(t\right)\beta_{\mathbf p}\left(t\right)\right] +2\Re\left[J_{x,\;{\mathbf p}}^{12}\left(t\right)\xi_{\mathbf p}^{*}\left(t\right)\gamma_{\mathbf p}\left(t\right)+J_{x,\;{\mathbf p}}^{23}\left(t\right)\gamma_{\mathbf p}^{*}\left(t\right)\beta_{\mathbf p}\left(t\right)\right].
\end{align}
\end{widetext}
To gain insights into these contributions to the average current density 
$\langle J_x \rangle_{\mathbf p}(t)$, we recall the matrix of the current density 
operator:
\begin{align} \nonumber
	J_{x,\;{\mathbf p}}\left(t\right)=-eU^{\dagger}_{\mathbf p}\left(t\right)\partial_{p_{x}(t)}H(t)U_{\mathbf p}\left(t\right).
\end{align}
The intraband contribution is made by both electrons and holes, corresponding to 
\begin{align} \nonumber
	J_{x,\;{\mathbf p}}^{11} &\left(t\right)\equiv J_{x,\;{\mathbf p}}^{0}\left(t\right)\cos\Theta_{\mathbf p}\left(t\right) \\ \nonumber
	J_{x,\;{\mathbf p}}^{33}& \left(t\right)=-J_{x,\;{\mathbf p}}^{11}\left(t\right), 
\end{align}
respectively. The interband contribution arises from the interference of the 
transitions from the lower to the flat band or the upper band and from the flat 
to the upper band, corresponding to 
$J_{x,\;{\mathbf p}}^{23}\left(t\right)$, $J_{x,\;{\mathbf p}}^{13}\left(t\right)$ and 
$J_{x,\;{\mathbf p}}^{12}\left(t\right)$, respectively, which are given by
\begin{align}
	J_{x,\;{\mathbf p}}^{13}\left(t\right)&\equiv i J_{x,\;{\mathbf p}}^{0}\left(t\right)\cos[2\varphi]\;\sin[\Theta_{\mathbf p}\left(t\right)],\nonumber\\
  J_{x,\;{\mathbf p}}^{12}\left(t\right)&\equiv i J_{x,\;{\mathbf p}}^{0}\left(t\right)\sin[2\varphi]\sin[\Theta_{\mathbf p}\left(t\right)]/\sqrt{2},\nonumber\\
  J_{x,\;{\mathbf p}}^{23}\left(t\right)&=J_{x,\;{\mathbf p}}^{12}\left(t\right),
\end{align}
where $\Theta_{\mathbf p}\left(t\right)\equiv\theta_{\mathbf p}\left(t\right)+Y_{p}$ 
and $J_{x,\;{\mathbf p}}^{0}(t)$ is the common factor with the dimension of the 
current density: 
\begin{align} \nonumber
J_{x,\;{\mathbf p}}^{0}\left(t\right)=-\sqrt{3}\sin [X_{\mathbf p}\left(t\right)]eat_{\epsilon}/\hbar.
\end{align}
We use $J(t)$ to denote the integration of $\langle J_x\rangle_{\mathbf p}(t)$ in the 
first Brillouin zone, which will be used to characterize the Bloch oscillations (in 
Sec.~\ref{subsec:Bloch_oscillation}).

For the general $\alpha$-$\mathcal{T}_{3}$ lattice calculations, the following
dimensionless quantities are convenient:
\begin{align}
    \widetilde{p}_x &= p_x/p_0,\nonumber\\
    \widetilde{p}_y &= p_y/p_0,\nonumber\\
    \widetilde{E} &=E/E_0,\nonumber\\
    \widetilde{J}(t) &=J(t)/J_0,\nonumber\\
    \langle \widetilde{J}_x\rangle_{\mathbf p}(t)
     &=\langle J_x\rangle_{\mathbf p}(t)/\langle J_{0}\rangle_{\mathbf p},
\end{align}
with $p_0\equiv \hbar/a$, $E_0 \equiv t_{\epsilon}/(ea)$, 
$J_0=e^{2}E_0/\hbar\sim et_{\epsilon}/(\hbar a)$ and 
$\langle J_0\rangle_{\mathbf p} = e a t_{\epsilon}/\hbar$. 

\subsection{Calculating the Berry phases of the $\alpha$-$\mathcal{T}_{3}$ lattice} \label{subsec:berry_phase_detection}

\begin{figure} [ht!]
\centering
\includegraphics[width=\linewidth]{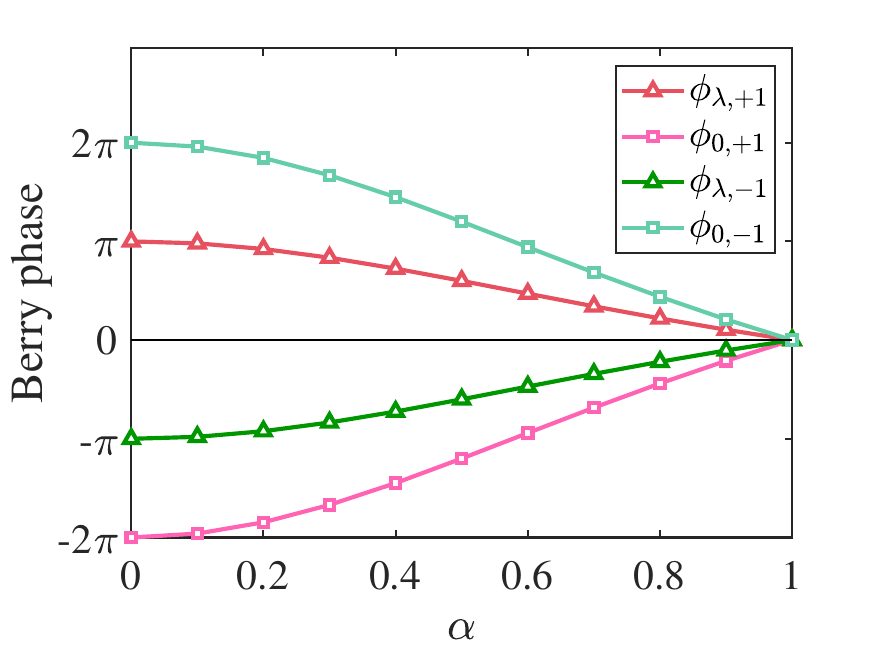}
\caption{Berry phases of orbits in different energy bands around the Dirac points 
$\pm \mathbf{K}$ versus the material parameter $\alpha$. The dependence of the Berry
phases on $\alpha$ is monotonic.}
\label{fig:berry_phase}
\end{figure}

The Berry phases associated with the conical and flat bands can be calculated by
assuming that the corresponding eigenstates adiabatically evolve with time along 
an arbitrarily closed loop around the Dirac points $\zeta {\mathbf K}$ in the momentum 
space~\cite{ICN:2015}:
\begin{equation}
\phi_{n,\zeta}=\frac{-i}{\pi}\oint dp\cdot \langle \psi_n|\mathbf{\nabla}_{\mathbf p}|\psi_n\rangle,
\end{equation}
which can be calculated either by the continuum effective Hamiltonian or by the
general lattice Hamiltonian (both giving the same results). For example, from the 
general lattice model, the eigenstate of the flat band is
\begin{align} \label{eq:flat_eigen} 
|\psi_{0}\rangle=\left[\begin{array}{c} \sin\varphi\;e^{i\theta_{\mathbf p}}\\
0 \\ -\cos\varphi\;e^{-i\theta_{\mathbf p}}
\end{array}\right],
\end{align}
and the eigenstates of the conduction and valence bands with $\lambda=\pm 1$, 
respectively, are
\begin{equation}\label{eq:conical_eigen}
|\psi_{\lambda}\rangle=\frac{1}{\sqrt{2}}\left[\begin{array}{c}
\cos\varphi\;e^{i\theta_{\mathbf p}}\\
\lambda\\
\sin\varphi\;e^{-i\theta_{\mathbf p}}
\end{array}\right],
\end{equation}
where $\theta_{\mathbf p}$ is the phase of the $f_{\mathbf p}$ in Eq.~\eqref{eq:f_p}. 
The eigenstates in the continuum effective model are similar to those in the general 
lattice model except that $\theta_{\mathbf p}$ is replaced by $\theta_{\mathbf q}$, 
the phase of the $f_{\mathbf q}$ in Eq.~\eqref{eq:f_q}. The Berry phases of the 
dispersive conical bands and the dispersionless flat band are given 
by~\cite{Raoux:2014,ICN:2015}
\begin{align}\label{eq:berry_phase}
    \phi_{\lambda,\;\zeta}&=\pi\zeta\cos2\varphi=\pi\zeta\left(\frac{1-\alpha^{2}}{1+\alpha^{2}}\right),\\
    \phi_{0,\;\zeta}&=-2\pi\zeta\cos2\varphi=-2\pi\zeta\left(\frac{1-\alpha^{2}}{1+\alpha^{2}}\right),
\end{align}
respectively. Note that the Berry phases are topological but not $\pi$ 
quantized~\cite{Raoux:2014} and are distinct in the $+\mathbf{K}$ and $-\mathbf{K}$ 
valleys except for $\alpha=0,1$. Figure~\ref{fig:berry_phase} shows that the Berry 
phases is a monotonic function of the material parameter $\alpha$. The average
current density in Eqs.~\eqref{eq:continuum_current density_1},
\eqref{eq:continuum_current density_2}, \eqref{eq:lattice_current density_1}
and \eqref{eq:lattice_current density_2} also depends on $\alpha$. If this 
dependence is monotonic, there will be a one-to-one correspondence between the 
current and the Berry phases, providing a mechanism to determine the Berry phases 
by measuring the current.

\section{Ballistic Transport and Berry Phase Detection} \label{sec:ballistic_transport}

\begin{figure*} [ht!]
\centering
\includegraphics[width=0.7\linewidth]{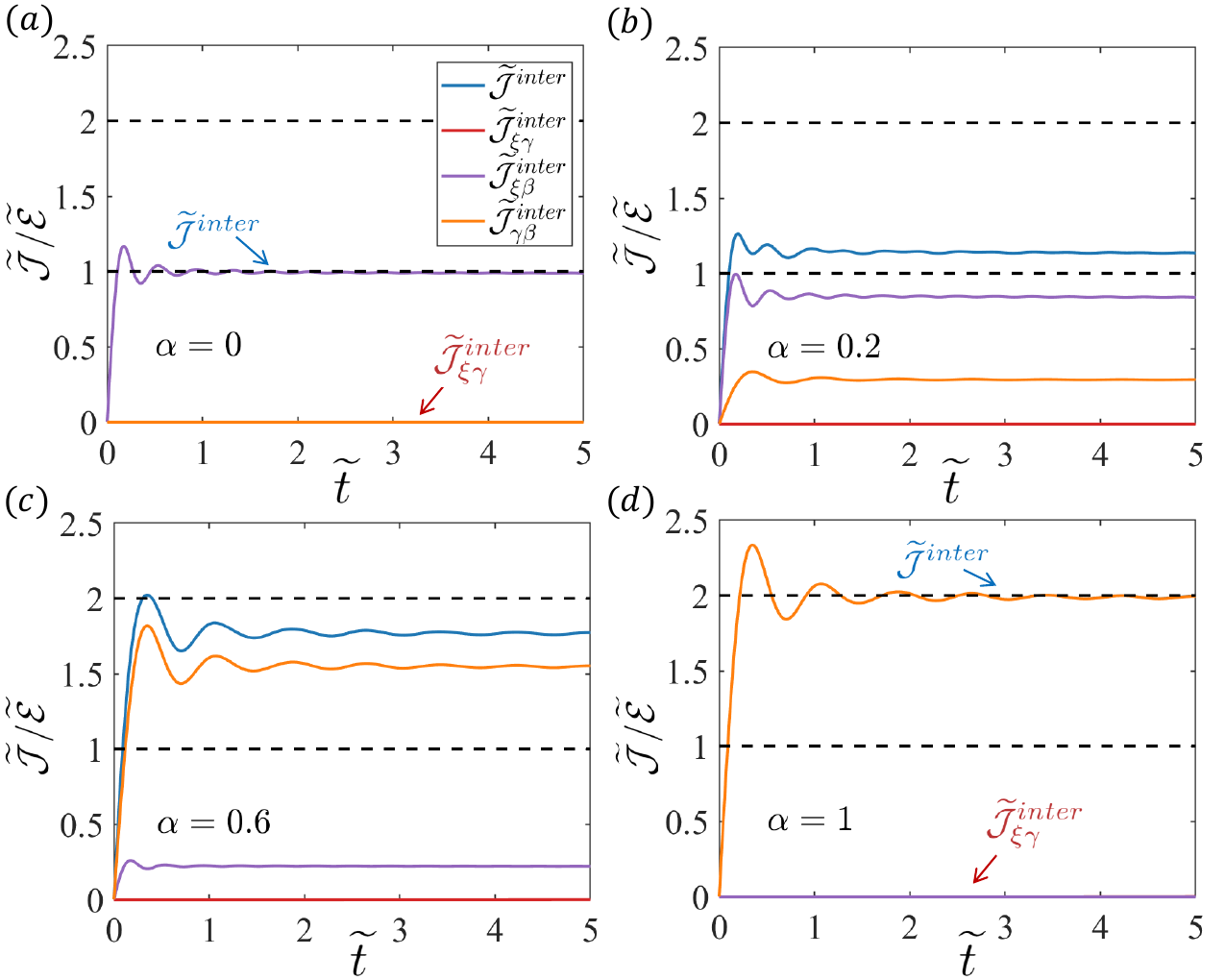}
\caption{Time evolution of the current from the effective continuum
Hamiltonian in the Kubo regime. Shown is the interband current divided by 
the electric field: $\widetilde{\mathcal{J}}/\widetilde{\mathcal{E}}$ for 
$0\leq\alpha\leq 1$. (a-d) Saturated currents divided by the electric field:
$\widetilde{\mathcal{J}}/\widetilde{\mathcal{E}}$, for the total 
interband current $\widetilde{\mathcal{J}}^{\textnormal{inter}}$, the 
interference current between the flat and the upper band 
$\widetilde{\mathcal{J}}^{\textnormal{inter}}_{\xi\gamma}$, the current between the lower and the upper band 
$\widetilde{\mathcal{J}}^{\textnormal{inter}}_{\xi\beta}$, and that between the 
lower and the flat band $\widetilde{\mathcal{J}}^{\textnormal{inter}}_{\gamma\beta}$ 
for $\alpha=0,\;0.2,\;0.6,\;1$, respectively. For comparison, all currents are
divided by $1/4$. Other parameters are: electric field 
$\widetilde{\mathcal{E}}=0.0004$, valley index of the Dirac point $\zeta=+1$, 
size of momentum space in $\widetilde{q}_{x},\widetilde{q}_{y}\in[-8,8]$ and  
step sizes of momentum and time $d\widetilde{q}=d\widetilde{t}=0.01$. Note the cut 
width about the Dirac point is $\widetilde{q}_{\textnormal{cut}}=0.005$ in the 
momentum space to make valid the weak field approximation and the quantum dynamic 
equation~\eqref{eq:dirac_q}.}
\label{fig:weak_t}
\end{figure*}

%{\color{Maroon}$$\widetilde{\mathcal{J}}^{inter}_{\xi\gamma}$$}
%{\color{Fuchsia}$$\widetilde{\mathcal{J}}^{inter}_{\xi\beta}$$}
%{\color{Maroon}$$\widetilde{\mathcal{J}}^{intra}_{\xi}$$}
\begin{figure} [ht!]
\centering
\includegraphics[width=\linewidth]{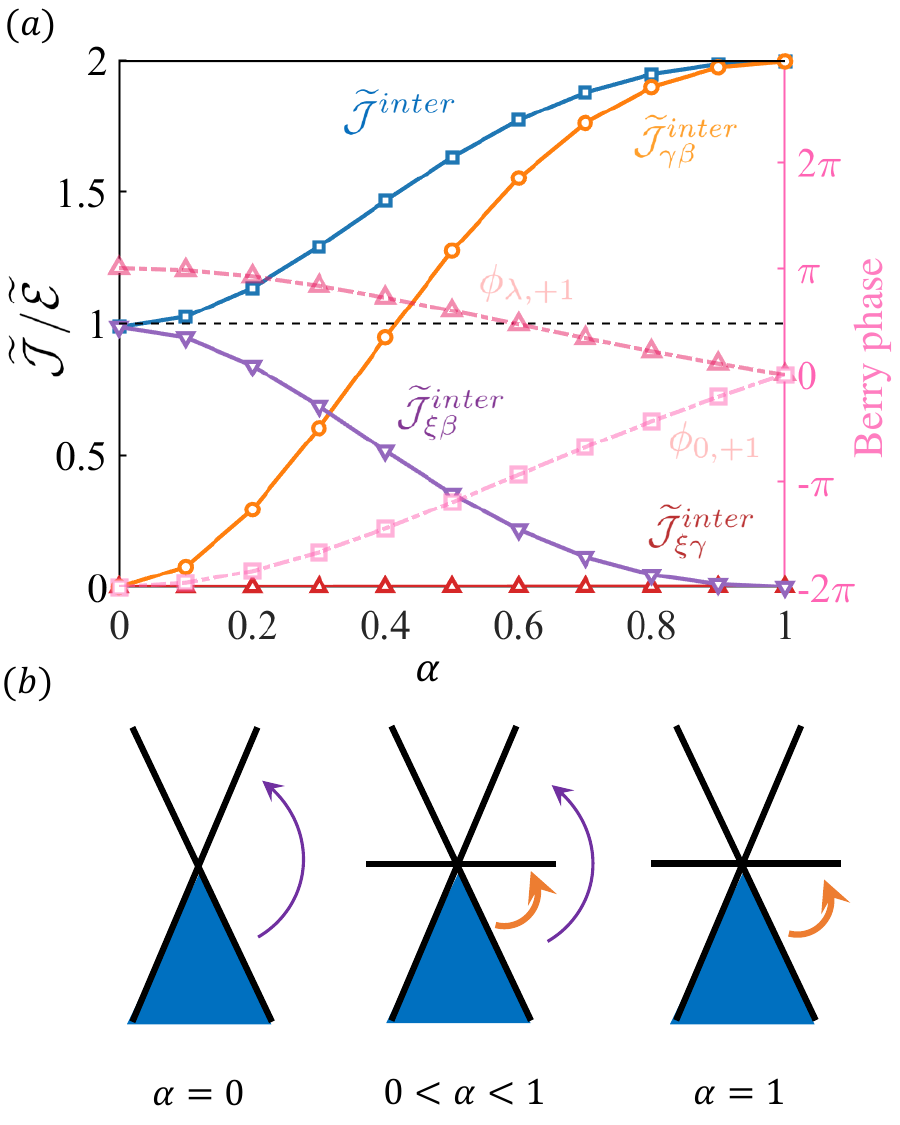}
\caption{Flat band contribution to the saturated current in the Kubo regime. 
(a) Corresponding to the Berry phase diagram in Fig.~\ref{fig:berry_phase}, the 
saturated current [the current at the end of time evolution in 
Fig.~\ref{fig:weak_t}] changes with the materials parameter $\alpha$ for the 
total interband contribution $\widetilde{\mathcal{J}}^{\textnormal{inter}}$, the 
interference current between the flat and upper bands 
$\widetilde{\mathcal{J}}^{\textnormal{inter}}_{\xi\gamma}$, the current 
between the lower and upper bands 
$\widetilde{\mathcal{J}}^{\textnormal{inter}}_{\xi\beta}$, and that between 
the lower and flat bands 
$\widetilde{\mathcal{J}}^{\textnormal{inter}}_{\gamma\beta}$.
(b) A schematic display of the mixing process of interference between the 
lower and upper bands and that between the lower and flat bands for 
$\alpha\in(0,1)$.}
\label{fig:weak_alpha_schematic}
\end{figure}

For nonequilibrium quantum transport in the $\alpha$-$\mathcal{T}_3$ lattice at 
zero temperature, depending on the time scale of ballistic transport, distinct 
physical behaviors can arise. First, in the presence of a uniform electric field, 
if its product with time is comparable to the quantity $\hbar/(ea)$: 
$Et_{\textnormal{Bloch}}\sim \hbar/(ea)$, the average current density will undergo 
Bloch oscillations~\cite{bloch:1928,zener:1934} due to the Bloch band in the periodic 
Brillouin zone. If, on this time scale, two levels do not cross each
other, Landau-Zener transition will occur, leading to Bloch-Zener 
oscillations~\cite{Ye:2023}. Second, if the time scale is much shorter than the Bloch 
time: $t\ll t_{\textnormal{Bloch}}$, the lattice can effectively be described by the 
continuum effective $\alpha$-$\mathcal{T}_3$ Hamiltonian. Third, when the time scale 
is in the Schwinger regime:
\begin{align} \nonumber
\sqrt{\hbar/(v_{F}e\mathcal{E})}\ll t\ll t_{\textnormal{Bloch}}, 
\end{align}
the transport process becomes nonlinear: $\mathcal{J}\propto t\mathcal{E}^{3/2}$. 
Fourth, when the time scale continues to reduce to the Kubo regime:
\begin{align} \nonumber
h/W\ll t\ll \sqrt{\hbar/(v_{F}e\mathcal{E})}, 
\end{align}
where $W$ is the bandwidth, the average electric current density is saturated and 
independent of time: $\mathcal{J}\propto \mathcal{E}$. Finally, for the ultrashort 
time transient response: $t\ll h/W$, the current behavior becomes full classical:
$\mathcal{J}\propto \mathcal{E}t$. 

To describe our results unambiguously, it is necessary to distinguish the 
electric field and current in the two cases where the $\alpha$-$\mathcal{T}_3$
material is described by the effective continuum model and by the general 
lattice model. We use $\mathcal{E}$ and $\mathcal{J}$ to denote the electric field 
and current in the former, while $E$ and $J$ in the latter.

\subsection{Kubo Regime} \label{subsec:weak_field_regime}

In the Kubo regime of the weak electric field, we have
\begin{equation}
	|{\mathbf q}|=\sqrt{q^{2}_{x}+q^{2}_{y}}\gg e\mathcal{E}t,
\end{equation}
for $|{\mathbf q}| \ne 0$ (not too close to the Dirac points). In 
Eq.~\eqref{eq:dirac_q}, the term $e\mathcal{E}t$ in $\varepsilon_{\mathbf q}(t)$ 
can then be neglected but the field term in the numerator term 
$-i\hbar U^{\dagger}_{\mathbf q}(t)\partial_{t}U_{\mathbf q}(t)$ should be
retained. Initially, at $t=0$, all electrons stay in the lower energy band. For 
$t>0$, a uniform constant electric field is switched on along the $x$ direction, 
and electron-positron pairs are created by the Schwinger 
mechanism~\cite{dora:2010,schwinger:1951} in the continuum effective model. 
Since only a small number of the particles are excited, the interband (or 
polarization) contribution from the interference between the energy bands dominates 
over the intraband (or conduction) contribution. In this regime, Zitterbewegung 
governs the small field linear response, where all electrons propagate with the 
maximal velocity $v_F$, leading to a saturated current independent of time.

Integrating the average current density associated with the momentum 
$\langle \mathcal{J}_{x}\rangle^{\textnormal{inter}}_{\mathbf q}(t)$ over the whole 
momentum space gives
\begin{align}
    \mathcal{J}^{\textnormal{inter}}&\equiv \langle \mathcal{J}_{x}\rangle^{\textnormal{inter}},\nonumber\\
    &=\frac{1}{\pi^2\hbar^2}\int^{\infty}_{0}q\;dq\int^{2\pi}_{0}d\varphi\;\langle \mathcal{J}_{x}\rangle^{\textnormal{inter}}_{\mathbf q}(t),
\end{align}
where $q$ and $\varphi$ are the radial and angular variables in momentum space, respectively. For pseudospin-$1/2$ quasiparticles ($\alpha = 0$), the linear scaling law for the current is~\cite{dora:2010} 
\begin{align} \nonumber
\mathcal{J}^{\textnormal{inter}}_{\textnormal{spin-1/2}} = e^2\mathcal{E}/(4\hbar) 
\end{align}
with the dimensionless relation 
\begin{equation}  
\mathcal{\widetilde{J}}^{\textnormal{inter}}_{\textnormal{spin-1/2}} = \frac{1}{4}\mathcal{\widetilde{E}}.
\end{equation}
For pseudospin-1 quasiparticles ($\alpha = 1$), due to the flat band, the current 
saturation value is amplified by a factor of two and the corresponding linear 
scaling law becomes~\cite{wang:2017}
\begin{equation}
\mathcal{\widetilde{J}}^{\textnormal{inter}}_{\textnormal{spin-1}} = \frac{1}{2}\mathcal{\widetilde{E}}.
\end{equation}
For $0<\alpha<1$, the saturated value of the current over the electric field 
$\widetilde{\mathcal{J}}^{\textnormal{inter}}/\widetilde{\mathcal{E}}$ is between 
$1/4$ and $1/2$, as shown in Fig.~\ref{fig:weak_t}, where the current is normalized
by the constant $1/4$. 

On the ultrashort time scale, the current exhibits a fully classical behavior: 
$\widetilde{\mathcal{J}}/\widetilde{\mathcal{E}}\propto \widetilde{t}$. After a certain
time, the current saturates. We tune the material parameter $\alpha$ to assess the 
interplay between the flat band and the saturated current in the weak field regime.
For $\alpha=0$, there is zero coupling between the flat and the two dispersive 
bands, so the interband current is solely determined by the interference between 
the lower and the upper conical bands as 
$\widetilde{\mathcal{J}}^{\textnormal{inter}}_{\xi\beta}$. For the opposite
extreme case of $\alpha=1$, the saturated current is the result of the interference 
between the lower and the flat band: 
$\widetilde{\mathcal{J}}^{\textnormal{inter}}_{\gamma\beta}$. For $0<\alpha<1$, the 
total interband saturated current is a mixture of the interference contributions
between $\widetilde{\mathcal{J}}^{\textnormal{inter}}_{\xi\beta}$ and 
$\widetilde{\mathcal{J}}^{\textnormal{inter}}_{\gamma\beta}$, as shown in 
Fig.~\ref{fig:weak_t}. Note that the interference between the flat and the upper 
bands does not directly contribute any current for the entire $\alpha$ spectrum, 
because the combination of the interference between the lower and flat bands and 
that between the flat and upper bands is physically equivalent to the interference
between the lower and upper bands.

As $\alpha$ increases from zero, the saturated current from the interference between 
the lower and upper bands decreases, and the current from the interference between
the lower and the flat band increases, as shown in 
Fig. \ref{fig:weak_alpha_schematic}(a). In the regime of weak field, the flat band 
suppresses the current from the interference between the lower and upper bands, 
and enhances the one from the interference between the lower and flat bands
for $\alpha\in(0,1)$, as shown in Fig.~\ref{fig:weak_alpha_schematic}(b). In this 
case, detecting the Berry phase through the current probe is feasible since there 
is a one-to-one correspondence between the saturated current value and the Berry 
phase, as shown in Figs.~\ref{fig:berry_phase} and \ref{fig:weak_alpha_schematic}(a).

\subsection{Schwinger Regime}\label{subsec:strong_field_regime}

\begin{figure*} [ht!]
\centering
\includegraphics[width=0.7\linewidth]{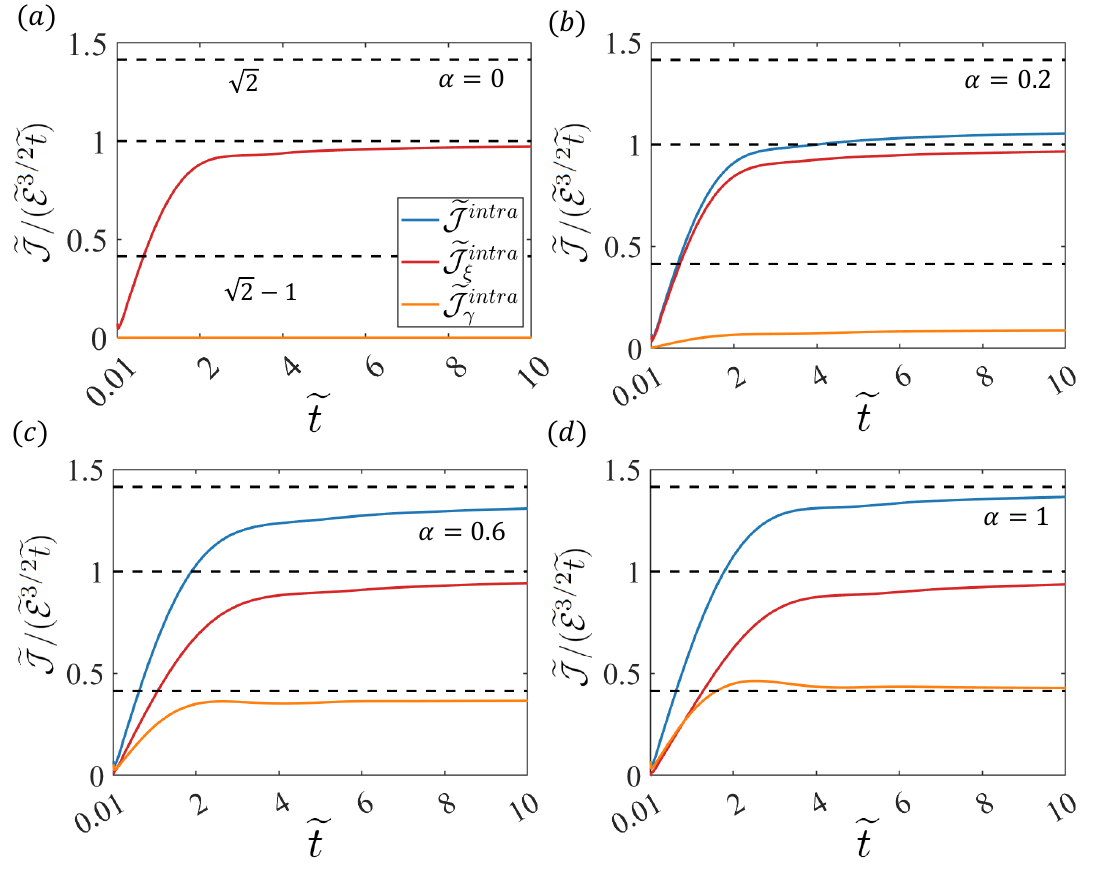}
\caption{Time evolution of the normalized intraband current from the continuum 
effective model in the Schwinger regime. The quantity displayed is 
$\widetilde{\mathcal{J}}/(\widetilde{\mathcal{E}}^{3/2}\widetilde{t})$. 
(a-d) The total intraband saturated current 
$\widetilde{\mathcal{J}}^{\textnormal{intra}}$ from the upper band current
$\widetilde{\mathcal{J}}^{\textnormal{intra}}_{\xi}$ and the flat band 
current $\widetilde{\mathcal{J}}^{\textnormal{intra}}_{\gamma}$ for 
$\alpha=0,\;0.2,\;0.6,\;1$, respectively. For comparison, all currents are
divided by $2/\pi^2$. Other parameters are: electric field 
$\widetilde{\mathcal{E}}=0.4096$, valley index $\zeta=+1$, size of momentum 
space $\widetilde{q}_{x},\widetilde{q}_{y}\in[-8,8]$, and step sizes of 
momentum and time $d\widetilde{q}=d\widetilde{t}=0.01$. The width of the region
in the momentum space about the Dirac point in which the continuum effective 
Hamiltonian holds is $\widetilde{q}_{\textnormal{cut}}=0.0001$.}
\label{fig:strong_t}
\end{figure*}

Under a strong electric field, 
\begin{equation}
    (q_x,eEt-q_x)\gg|q_y|,
\end{equation}
the $\alpha$-$\mathcal{T}_3$ lattice system is in the Schwinger regime. In this
regime, electrons are excited from the lower band to the flat and upper bands
via the Schwinger mechanism, where the electric field in the vacuum decays and loses 
energy due to the production of the electron-positron pairs. The transition 
probability to the flat or upper band is the same as the Landau-Zener 
transition probability, where the finite energy gap between the two avoided-crossing 
levels induces the nonadiabatic Landau-Zener transition driven by the electric field.
In the general $\alpha$-$\mathcal{T}_3$ lattice or the corresponding continuum model,
Landau-Zener transitions occur in the neighborhood of the Dirac points~\cite{Ye:2023} 
because the energy gap is comparable with that given by the magnitude of the 
electric field. After the Landau-Zener transition, the quantum states become superposition states between three levels~\cite{Ye:2023}, and the nonzero component of occupied probability in the flat band reveals the Landau-Zener transition from the lower band to the flat band. This transition leaves more holes in the lower band, thereby increasing the current. 

Specifically, in the Schwinger regime, the electric current is dominated by the 
intraband transitions, including the flat band contribution. The flat band is defined as being dispersionless because it has zero group velocity for wave packets. This means that wave packets corresponding to the flat band are localized in real space. However, the flat band also contributes to the intraband current even though its group velocity is zero, for the following reasons. First, both electrons 
and holes contribute to the current. Second, the Landau-Zener transition from the lower 
to the flat band can create relatively more holes, giving rise to an extra current 
compared with the case without a flat band. As a result, the intraband current is 
proportional to the number of excited particles in both the flat and upper bands.

\begin{figure} [ht!]
\centering
\includegraphics[width=\linewidth]{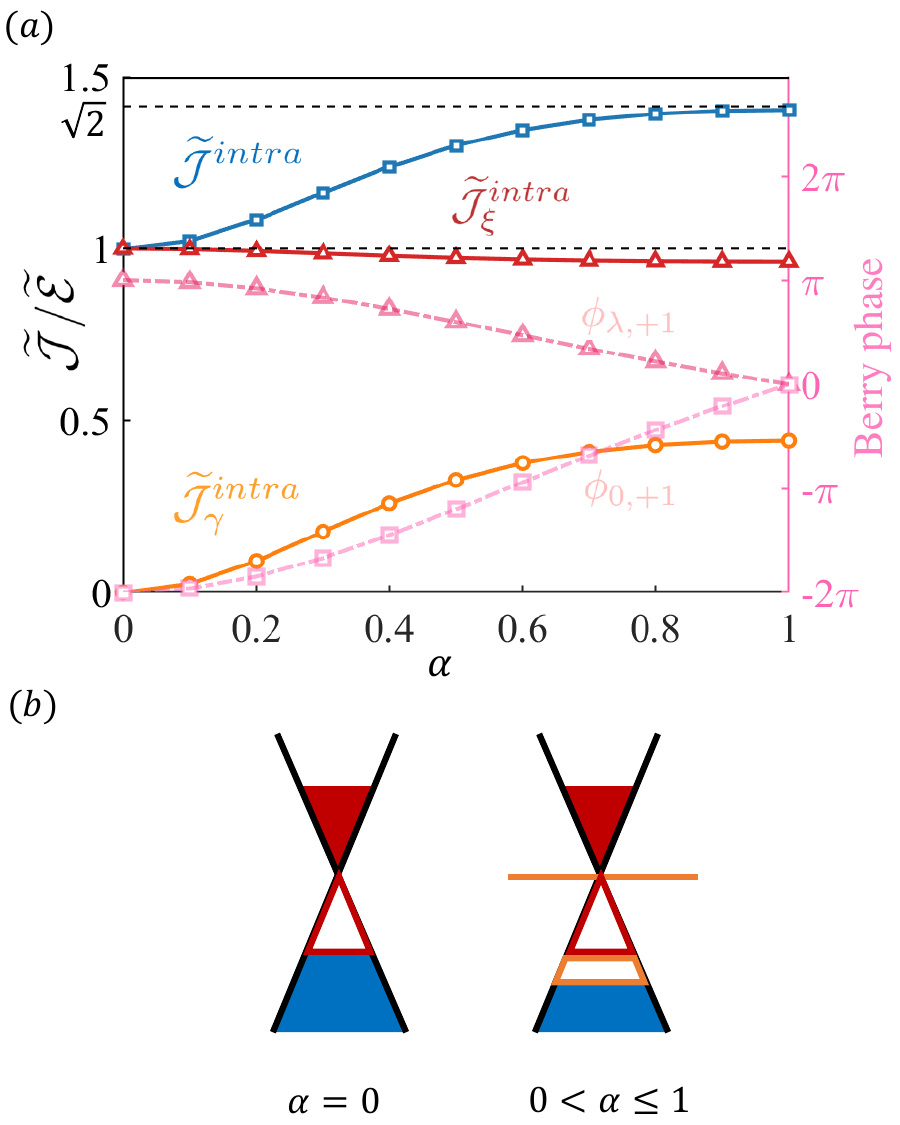}
\caption{Flat band contributions to the saturated current in the Schwinger regime.
(a) Corresponding to the Berry phase diagram in Fig.~\ref{fig:berry_phase}, the 
saturated current (at the end of the time evolution in Fig.~\ref{fig:strong_t}) varies 
with the material parameter $\alpha$ for the total intraband current density 
$\widetilde{\mathcal{J}}^{\textnormal{intra}}$, the upper band current
$\widetilde{\mathcal{J}}^{\textnormal{intra}}_{\xi}$, and the flat 
band current $\widetilde{\mathcal{J}}^{\textnormal{intra}}_{\gamma}$ for 
$0 \le \alpha \le 1$. (b) A schematic illustration of the electron-hole excitation 
and the extra holes from the flat band in comparison with the graphene case. The flat 
band, despite its zero group velocity, has the ability to enhance the electric 
current.}
\label{fig:strong_alpha}
\end{figure}

Integrating the current 
$\langle\mathcal{J}_{x}\rangle^{\textnormal{intra}}_{\mathbf q}(t)$ 
over the whole momentum space gives 
\begin{align} \label{eq:intergration_J}
    \mathcal{J}^{\textnormal{intra}}(t)&\equiv \langle \mathcal{J}_{x}\rangle^{\textnormal{intra}}(t),\nonumber\\
    &=\frac{1}{\pi^2\hbar^2}\int^{\infty}_{0}q\;dq\int^{2\pi}_{0}d\varphi\;\langle \mathcal{J}_{x}\rangle^{\textnormal{intra}}_{\mathbf q}(t),
\end{align}
with $q$ and $\varphi$ being the radial and angular variables in momentum space, respectively. This form is similar in mathematical form to that of the interband case. However, a key difference is that the time-dependent intraband current displays a nonlinear 
response. For pseudospin-1/2 Dirac particles, the dimensionless form of the 
intraband current is given by~\cite{dora:2010}
\begin{equation}\label{eq:spin_half_intraband}
    \widetilde{\mathcal{J}}^{\textnormal{intra}}_{\textnormal{spin-1/2}}(\widetilde{t}) = \frac{2}{\pi^2} \widetilde{\mathcal{E}}^{3/2}\widetilde{t}.
\end{equation}
For pseudospin-1 Dirac-Weyl particles, the intraband current is~\cite{wang:2017}
\begin{equation}\label{eq:spin_one_intraband}
    \widetilde{\mathcal{J}}^{\textnormal{intra}}_{\textnormal{spin-1}}(\widetilde{t}) = \frac{2\sqrt{2}}{\pi^2} \widetilde{\mathcal{E}}^{3/2}\widetilde{t},
\end{equation}
where the flat band contribution is
\begin{equation}
    \widetilde{\mathcal{J}}^{\textnormal{intra}}_{\textnormal{flat}}(\widetilde{t}) = \frac{2(\sqrt{2}-1)}{\pi^2} \widetilde{\mathcal{E}}^{3/2}\widetilde{t}.
\end{equation}
The continuum effective $\alpha$-$\mathcal{T}_3$ Hamiltonian can be used to gain 
insights into the origin of the intraband current in the Schwinger regime. For 
$\alpha=0$ with pseudospin-1/2 Dirac particles, there is no coupling between the
flat band and the two conical dispersive bands, so the only excitation is one from 
the lower to the upper band. The intraband current depends only on the Landau-Zener 
transition to the upper band, as illustrated in Fig.~\ref{fig:strong_t}(a). 

Graphene can serve as a benchmark for comparison with the general $\alpha > 0$ cases. 
For the opposite extreme case of $\alpha=1$ with pseudospin-1 Dirac-Weyl particles, 
the quantity $\widetilde{\mathcal{J}}/(\widetilde{\mathcal{E}}^{3/2}\widetilde{t})$ 
of the upper band is in principle the same as that for the $\alpha=0$ case, with the 
current from the Landau-Zener transition to the flat band converging to the 
constant $\sqrt{2}-1$, as shown in Fig.~\ref{fig:strong_t}(d). For $0<\alpha< 1$, 
the intraband current of the upper band is approximately constant and flat-band
current is enhanced with increasing $\alpha$, as shown in 
Figs.~\ref{fig:strong_t} (a-d) and Fig.~\ref{fig:strong_alpha} (a). In the Schwinger regime, the flat band contributes extra holes with positive charges, thereby 
enhancing the intraband current by the factor of $\sqrt{2}$ compared with the 
graphene benchmark, as schematically illustrated in Fig.~\ref{fig:strong_alpha}(b). 
Since the converged intraband current depends monotonically on the materials
parameter $\alpha$, it can be exploited to assess the Berry phase.

In the Schwinger regime, the total difference of the intraband current in the prefactor between pseudospin-1/2 and pseudospin-1 Dirac-Weyl particles from Eq.~\eqref{eq:spin_half_intraband} and Eq.~\eqref{eq:spin_one_intraband}, respectively, physically arises from intrinsic dynamics, specifically the Landau-Zener transition in the Schwinger regime.  The final results in Eq.~\eqref{eq:spin_half_intraband} and Eq.~\eqref{eq:spin_one_intraband} have a prefactor difference of $\sqrt{2}$. The spin-1/2 and spin-1 matrices also differ by a prefactor of $1/\sqrt{2}$. Note that spin-1/2 matrices are two-by-two matrices, while spin-1 matrices are three-by-three. There is no evidence to show that the prefactor difference in the current physically originates from the prefactor difference of spin-1/2 and spin-1 matrices. Mathematically, in the effective model, the intraband current $\mathcal{J}^{\textnormal{intra}}(t)$ is, in principle, the integration of the average intraband current density over the infinite momentum space $\langle\mathcal{J}_{x}\rangle^{\textnormal{intra}}_{\mathbf q}(t)$ from Eq.~\eqref{eq:intergration_J}. In the Schwinger regime, the intraband contribution dominates, so the interband current effect is neglected. The total average current density $\langle \mathcal{J}_{x}\rangle_{\mathbf q}\left(t\right)$ is defined from the original Hamiltonian in Eq.~\eqref{eq:current_original}. From this formula, we see that the current contribution depends not only on the spin matrix $S'_x(\varphi)$ but also on the intrinsic time dynamics evolution of the wave function $\psi_{\mathbf{q}}(t)$. Furthermore, the derived intraband average current density is given by Eq.~\eqref{eq:continuum_current density_1}, which also does not provide explicit evidence that the current depends solely on the spin matrices.

\begin{figure} [ht!]
\centering
\includegraphics[width=\linewidth]{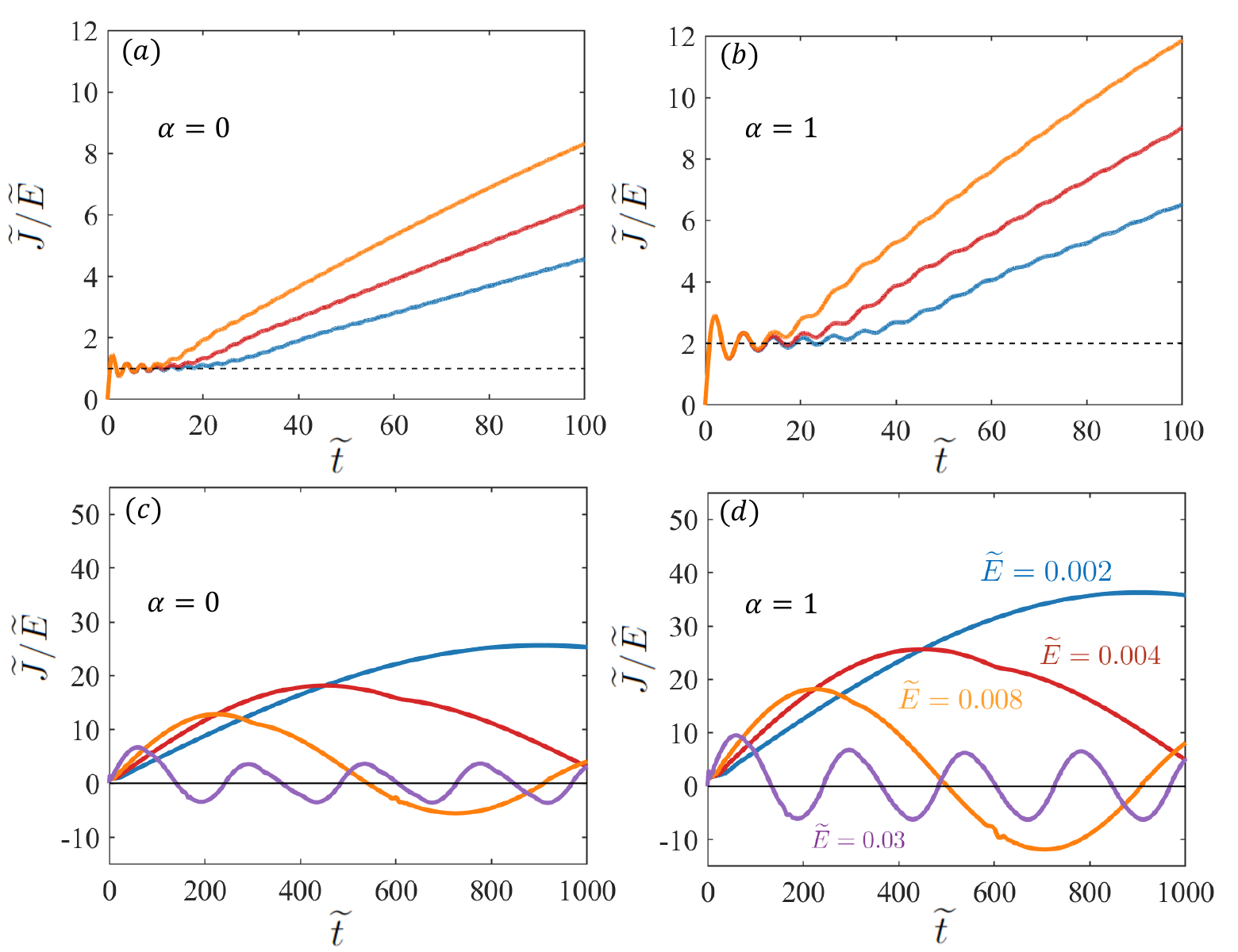}
\caption{Time evolution of the total current (interband and intraband) calculated
from the general $\alpha$-$\mathcal{T}_3$ lattice model. The quantity displayed is
$\widetilde{J}/\widetilde{E}$. (a, b) Ultrashort time transient response as well as 
the linear and nonlinear response for $\widetilde{E}=0.002,\;0.004,\;0.008$ and for 
$\alpha=0,\;1$, respectively. For comparison, all currents are divided by $1/4$, 
the saturated value of the weak field for graphene, and a factor of two due to the
momentum integration region being the first hexagonal Brillouin zone that contains 
two nonequivalent Dirac points. (c, d) Bloch-Zener oscillations for $\alpha=0,\;1$,
respectively, for different electric fields. Relevant parameter values are: time 
step size $d\widetilde{t}=0.01$, momentum step size 
$d\widetilde{p}_{x}=d\widetilde{p}_{y}=0.002$, width around the Dirac point 
$\widetilde{p}_{\textnormal{cut}}=0.001$ in the momentum space.}
\label{fig: Bloch_oscillation_t}
\end{figure}

\begin{figure} [ht!]
\centering
\includegraphics[width=\linewidth]{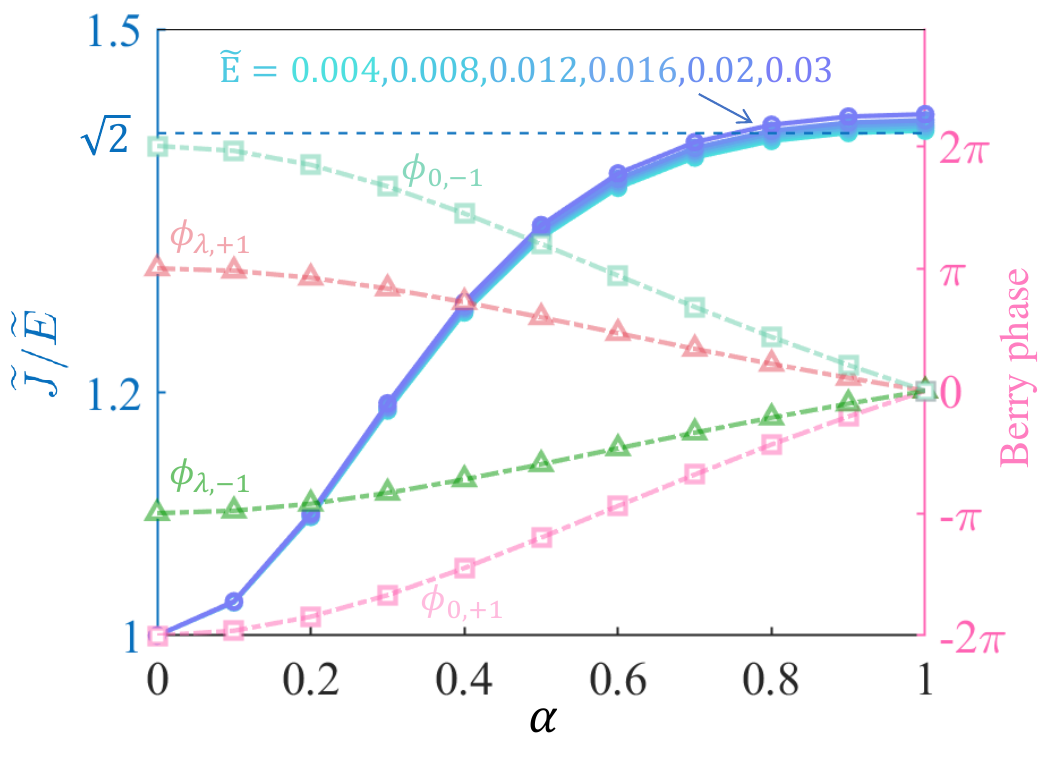}
\caption{Scaling law of the first peak of Bloch-Zener oscillations with the 
material parameter $\alpha$. Shown are the scaling relations for different values
of the electric field corresponding to the Berry-phase plots in 
Fig.~\ref{fig:berry_phase} for quantum states from the conical and flat bands 
around the Dirac points $\pm K$.}
\label{fig: Bloch_oscillation_alpha_berry_phase}
\end{figure}

\begin{figure*} [ht!]
\centering
\includegraphics[width=0.8\linewidth]{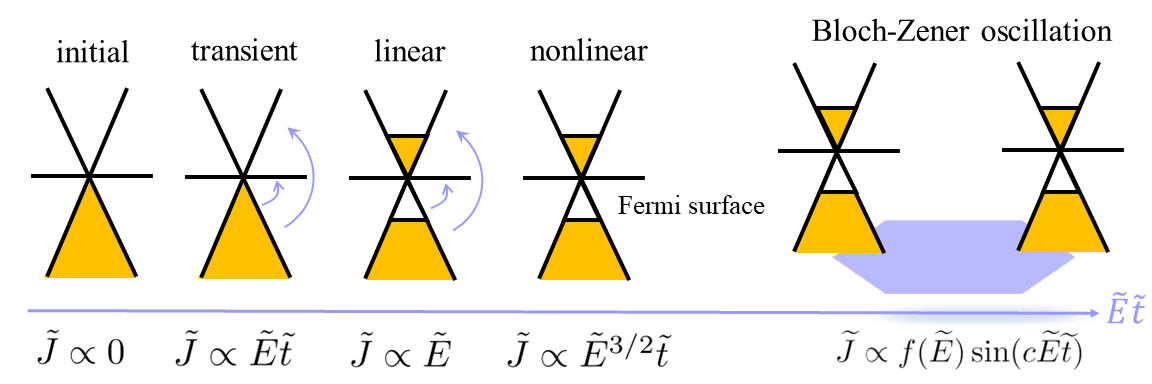}
\caption{Summary diagram: scaling behaviors, transitions, and physical mechanisms 
associated with ballistic transport in $\alpha$-$\mathcal{T}_{3}$ lattice on 
different time scales.}
\label{fig:summary}
\end{figure*}

\subsection{Bloch-Zener oscillations} \label{subsec:Bloch_oscillation}

The linear and nonlinear responses are obtained from the continuum effective 
$\alpha$-$\mathcal{T}_{3}$ model that is valid for low-energy excitations. More
experimentally relevant is the general lattice model. Here, using this model, we
calculate the currents for the Kubo, Schwinger, and Bloch-Zener oscillation regimes. 
Figures~\ref{fig: Bloch_oscillation_t}(a) and \ref{fig: Bloch_oscillation_t}(b) 
show that the ultrashort time transient, linear and nonlinear responses generated
by the continuum effective model persist for the lattice model with the respective
time-scale increment factors $h/W$, $\sqrt{{\hbar/(v_{F}e\mathcal{E})}}$,
$t_{\textnormal{Bloch}}\sim\hbar/(eaE)$, for any fixed electric field. In the Kubo 
regime, the total current $\widetilde{J}/\widetilde{E}$ still saturates. The
consistency between Figs.~\ref{fig: Bloch_oscillation_t}(a-b) and 
Fig.~\ref{fig:weak_alpha_schematic}(a) suggests that the linear response can be 
used to detect the Berry phase. In the Schwinger regime, the current is the result 
of nonlinear response  
\begin{equation}
    \widetilde{J}/\widetilde{E}\propto \widetilde{E}^{1/2}\widetilde{t},
\end{equation}
as shown in Figs.~\ref{fig: Bloch_oscillation_t}(a-b).

In the Kubo regime where the time scale is larger than the classical time $h/W$,
the interference between energy bands begins to contribute to the Landau-Zener 
transitions. In the Schwinger regime, the nonlinear response is dominated by 
the Landau-Zener transitions. For $Et_{\textnormal{Bloch}}\sim \hbar/(ea)$, Bloch 
oscillations occur. The combination of the Landau-Zener transition and Bloch 
oscillation leads to Bloch-Zener oscillations, as shown in 
Fig.~\ref{fig: Bloch_oscillation_t}(c-d) for $\alpha=0,\;1$, where the Bloch 
time period is $\widetilde{t}_{B}=4\pi/(\sqrt{3}\widetilde{E})$. The decay of the 
amplitude and the irregular behavior of the Bloch-Zener oscillations in 
$\alpha$-$\mathcal{T}_{3}$ lattice are the result of mixed interference of quantum 
states in multiple bands modulated by the geometric and dynamic phases~\cite{Ye:2023}.
For a range of the electric field, the time scales of the ultrashort transient 
and linear responses can be neglected compared with that of the nonlinear response. 
In this case, the first peak in the Bloch-Zener oscillations displays a scaling law,
as shown in Fig.~\ref{fig: Bloch_oscillation_alpha_berry_phase}. 

Taken together, the first peak in the Bloch-Zener oscillations, the nonlinear 
response, and the saturated current associated with the linear response 
all depend monotonically on the material parameter $\alpha$. These physical quantities
can then be exploited to detect the Berry phase.

\section{Discussion} \label{sec:discussion}

The Berry phase in the $\alpha$-$\mathcal{T}_{3}$ lattice varies monotonically with 
the material parameter $\alpha$. We investigated electronic transport when an 
$\alpha$-$\mathcal{T}_{3}$ lattice system is driven by a constant electric field 
and calculated a number of current densities as a function of $\alpha$ in both the 
linear (Kubo) and nonlinear (Schwinger) response regimes. Remarkably, the current 
density also exhibits a monotonic dependence on $\alpha$, implying that the Berry
phase as a fundamental material characteristic can be determined by measuring the 
current (e.g., using graphene for calibration). 

The various experimentally relevant 
scaling behaviors of the current density concerning the electric field and time 
as well as the underlying state transitions are summarized in Fig.~\ref{fig:summary}. 
Depending on the product $\tilde{E}\tilde{t}$ of the normalized electric field and 
time, five distinct scaling regimes arise. For $\tilde{E}\tilde{t} \sim 0$, the 
current density is zero. In the transient phase, the current density is proportional 
to $\tilde{E}\tilde{t}$. The linear response regime comes after the transient phase, 
in which the current is proportional just to the electric field. In the nonlinear 
response regime that follows the linear regime, the current is proportional to 
$\tilde{E}^{3/2}\tilde{t}$. For much larger values of $\tilde{E}\tilde{t}$, 
Bloch-Zener oscillations arise, whose amplitude can be an irregular function of 
time~\cite{Ye:2023}. While the scenario in Fig.~\ref{fig:summary} is based on the 
effective continuum Hamiltonian, direct calculations of the lattice Hamiltonian 
indicate that the ultrashort time transient response, linear and nonlinear responses 
still arise. In fact, Landau-Zener transitions begin to occur in the linear response 
regime and become dominant in the nonlinear response regime. When 
$\tilde{E}\tilde{t}$ is comparable to a quantity of the same physical dimension 
determined by the lattice constant, Bloch-Zener oscillations occur. In this case, the 
first peak of the oscillation exhibits a scaling law with $\alpha$ by the nonlinear 
response mechanism when the ultrashort time transient and linear responses are 
negligible. Consequently, the linear and nonlinear responses can be exploited for 
experimental detection of the Berry phase, so can the Bloch-Zener oscillations, 
as the time scale around the time for the first oscillation peak to occur is 
currently experimentally feasible~\cite{rosenstein:2010}.

Zitterbewegung oscillations are not exclusive to Dirac electrons~\cite{rusin:2008}. They also occur experimentally in various physical systems characterized by linear dispersion, such as ultracold atoms~\cite{vaishnav:2008}, a photonic crystal~\cite{zhang:2007}, a Bose-Einstein condensate~\cite{qu:2013,leblanc:2013}, a photonic microcavities~\cite{lovett:2023}, etc. Zitterbewegung provides physical interpretation for minimal conductivity in graphene~\cite{katsnelson:2006} and conductance fluctuations in quantum wells~\cite{iwasaki:2017}, and offers a calculation method for optical conductivity~\cite{oriekhov:2022,oriekhov:2021}. These oscillations can be interpreted as a measurable consequence of the momentum-space Berry phase, as the amplitude of Zitterbewegung can be modulated by the Berry phase~\cite{vaishnav:2008,cserti:2010,oriekhov:2022,biswas:2018}. It is believed that Zitterbewegung occurs due to the interference between positive and negative energy solutions of the Dirac equation~\cite{biswas:2018} in the effective low-energy approximation model.  So far, there exists only one study~\cite{biswas:2018} on the zitterbewegung effect in the alpha-T3 effective model. In our work, we calculate the linear response in quantum transport and reveal that the interplay between Zitterbewegung and Berry phases exists in both the lattice model and the low-energy effective model in an $\alpha$-$\mathcal{T}_{3}$ lattice. 

Theoretically, it is possible to develop a semiclassical correspondence with our results, provided the prerequisites for the WKB approximation are met. These prerequisites include a potential that varies slowly compared to the wavelength of the particle, energy levels not being too close to the turning points, and consideration of the Klein tunneling effect from quantum relativistic nature, among others. The previous work~\cite{weekes:2021} is a leading study in the semiclassical treatment of Dirac electrons. Under a finite mass potential, they address the effective $\alpha$-$\mathcal{T}_3$ model, utilizing the Wentzel-Kramers-Brillouin (WKB) semiclassical method to investigate the transmission of Dirac electrons. The finite mass potential opens a gap in the energy spectrum. However, our model features a zero energy gap due to the absence of a mass potential. We incorporate a time-dependent vector potential into the momentum to study the interaction of massless Dirac electrons with an electric field. Even so, with the necessary modifications based on the previous work~\cite{weekes:2021}, our research still holds promise for establishing a semiclassical correspondence.

\section*{Acknowledgment}

This work was supported by AFOSR under Grant No.~FA9550-21-1-0186.

\bibliography{Berry_Phase}

\end{document}